\documentclass[epsfig,bm,showpacs, twocolumn, preprintnumbers]{revtex4}
\usepackage{amsmath, amsfonts,euscript,bbm}
\usepackage{ifthen}
\usepackage{graphicx}
\usepackage{psfrag}

\newboolean{Notes}\setboolean{Notes}{true}
\newcommand{\notes}[1]
        {\ifthenelse{\boolean{Notes}}{{\tt #1}}{}}
        
\newcommand{\bas}{\begin{eqnarray*}}
\newcommand{\eas}{\end{eqnarray*}}
\newcommand{\ba}{\begin{eqnarray}}
\newcommand{\ea}{\end{eqnarray}}

\def\r{r_{\rm{CMB}}}
\def\T{\mathcal{S}}
\def\NE{\mathcal{N}_{\rm{end}}}
\def\Ne{\mathcal{N}_{\rm{eff}}}
\def\N{\mathcal{N}}

\def\ap{\alpha^{\prime}}
\def\ba{\begin{eqnarray}}
\def\ea{\end{eqnarray}}
\def\be{\begin{equation}}
\def\ee{\end{equation}}
\def\bas{\begin{eqnarray*}}
\def\eas{\end{eqnarray*}}

\def\det{\text{det}}

\def\t{\theta}
\def\w{\wedge}

\def\hf{\frac12}
\def\ap{{\alpha^{\prime}}}

\def\N{\mathcal N}

\def\N{\mathcal{N}}

\begin{document}

\title{Inflation from Wrapped Branes}
\author{Melanie Becker$^1$}
\email{mbecker@physics.tamu.edu}
\author{Louis Leblond$^1$}
\email{lleblond@physics.tamu.edu}
\author{Sarah Shandera$^2$}
\email{sarah@phys.columbia.edu}
 \affiliation{  $^1$ George P. \& Cynthia W. Mitchell Institute for Fundamental Physics,
  Texas A\&M University, College Station, TX 77843-4242\\
$^2$Institute of Strings, Cosmology and Astroparticle Physics,
Physics Department, Columbia University, New York, NY 10027}

\begin{abstract}
We show that the use of higher dimensional 
wrapped branes can significantly extend the inflaton field range compared to brane inflation models which use $D$3-branes. 
We construct a simple inflationary model in terms of
5-branes wrapping a $2$-cycle and traveling
towards the tip of the Klebanov-Strassler throat. Inflation ends
when the branes reach the tip of the cone and self-annihilate. 
Assuming a quadratic potential for the brane it is possible to match the CMB data in the DBI regime, but
we argue that the backreaction of the brane is important and cannot be neglected. 
This scenario predicts a strong non-Gaussian signal and possibly detectable 
gravitational waves.
\end{abstract}
\preprint{MIFP-07-23}
\maketitle

\section{Introduction}
Despite much recent progress, completely successful models of
inflation in string theory are still hard to come by. Many models
that attempt to achieve standard, single field, slow-roll inflation
suffer from fine-tuning issues associated with the necessary
flatness of the potential. In addition, these models often do not have any particularly stringy features and so, while encouraging, are not very useful as tests of string theory. In this paper we are interested in
the brane inflationary scenario \cite{Dvali:1998pa,
Burgess:2001fx, Alexander:2001ks, Jones:2002cv} in which the
inflaton is described by the position of a brane. Because of the unusual form of the brane action, this scenario can provide enough inflation even with a steep potential and has potentially ``stringy" signatures such as observably large non-Gaussianity and cosmic strings. 

The most studied case is $D$3-$\bar D$3 brane inflation in a warped
geometry \cite{Kachru:2003sx}.  In this scenario the inflaton is the
distance between the branes and inflation ends when the branes
annihilate at the bottom of the throat. While the warped geometry
helps to flatten the Coulombic potential between the branes, effects from compactification and interactions with other branes or fluxes used to stabilize moduli contribute to the potential. These terms must be fine-tuned to achieve slow-roll inflation.  Of course, one can argue that with
 a large landscape of string theory vacua, there will be one where the necessary
 flat potential is achieved in a self-consistent regime.  This has been checked in detail
 in \cite{Krause:2007jk, Baumann:2007np, Baumann:2007ah} which found that it is sometimes possible 
 to achieve slow-roll inflation in the $D$3 system, albeit of a particular type.

However, the square root structure of the brane kinetic term can play an interesting role in this type of model \cite{Silverstein:2003hf, Alishahiha:2004eh}. The square root imposes a speed limit on the canonical inflaton, $\phi$, so that even in a steep potential enough e-folds may be generated to solve the usual horizon problem. The form of the action is
\be
\label{act}
S=-\int d^4x\;a(t)^3 f(\phi)^{-1}\sqrt{1-f(\phi)\dot{\phi}^2} +V(\phi)\; ,
\ee
where $a(t)$ is the scale factor and $V(\phi)$ is the inflaton potential. The function $f(\phi)$ causes a field-dependent speed limit for the inflaton and
can be chosen such that the inflaton moves slowly (but relativistically in the sense that its
velocity approaches the local limit) even in a steep potential. In the limit of small proper velocity, this kinetic term reduces to the usual one. When the brane is moving in a regime where the square root cannot be expanded, the scenario has been called DBI (Dirac-Born-Infeld) inflation. A Lorentz factor for the inflaton, $\gamma(\phi) = \frac{1}{\sqrt{1-f(\phi)\dot\phi^2}}$, captures the differences between this theory and one with a usual kinetic term. The sound speed in this model is equal to $1/\gamma$, which changes the time of horizon exit for scalar modes. As a result, the tensor-scalar ratio $r$ is suppressed by $\gamma$, so that for a changing speed limit (changing sound speed) $r$ may be rapidly suppressed. The model also comes with large non-Gaussianities (ideally an observational blessing) which are severely constrained by
current observations, making it difficult to fit all data
consistently. In fact, Baumann and McAllister \cite{Baumann:2006cd} 
have recently shown that DBI inflation with
$D$3-branes in a throat can hardly match the data in a regime where
the supergravity is self-consistent. We will
review this fact here and demonstrate that the constraints are less stringent for wrapped branes. 

One of the strongest microscopic constraints on brane inflation has
always been the field range, i.e. the fact that the inflaton can
only move a very limited distance. This is easy to see from the fact
that the moduli space of the brane position is just the physical
compactification space itself. This needs to be compact for a
realistic phenomenology with gravity.
 The field range $\Delta \phi/M_p$ is constrained by the relationship between the four-dimensional
 (reduced) Planck scale $M_p$ and the six-dimensional volume. In the old slow-roll brane
 inflation on a torus, this limited field range meant that it was impossible to get enough e-folds
 \cite{Burgess:2001fx}. For $D$3-brane inflation in a warped throat with a quadratic potential, a typical field range that
 gives the observed COBE normalization comes into conflict with the current bound
 on non-Gaussianity \cite{Baumann:2006cd}. Increasing the field range without violating this bound
 is then the key to
 matching observations. As we will show in this note, this can be done by describing the inflaton
 of KKLMMT \cite{Kachru:2003sx} in terms of the position of branes wrapped on cycles
  of the internal geometry.
Furthermore, taking into account
 the feature that the sound speed may vary considerably
 during inflation, one may potentially obtain an observable value of the tensor-scalar
 ratio.  On the downside, we find that higher dimensional wrapped branes have a larger
 backreaction and it seems impossible to have a valid probe brane analysis throughout the 
 inflationary era. However, it is not clear that this prohibitively alters the observable quantities. 

Before going into detail, we first note that DBI inflation with a quadratic potential is just chaotic inflation with small sound speed (although it is hybrid in the sense that a tachyonic field kicks in to end inflation). If the Hubble parameter is not dominated by a constant, so $H(\phi)=h_n\phi^n$, $n>0$, then to have inflation, 
\ba
\label{infcondition}
\epsilon&=&\frac{2M_p^2}{\gamma}\left(\frac{H^{\prime}}{H}\right)^2<1\; ,\\\nonumber
&\Rightarrow&\frac{\phi}{M_p}>n\sqrt{\frac{2}{\gamma}}\; ,
\ea
where $M_p$ is the reduced Planck mass, the derivative of $H$ is with respect to the inflaton field and
the varying sound speed is included through the $\gamma$ factor. So trans-Planckian field range is required unless $\gamma$ is large (sound speed is small). In DBI inflation, $\gamma$ is constrained by data to be not too large and the field range is restricted geometrically. So while DBI inflation brings the necessary field range below the Planck scale, it only does so barely. In understanding the conflict with data, it is useful to keep that simple idea in mind.

The rest of the paper is divided as follows. We first review the status of brane inflation and discuss how to
generalize the familiar $D$3 scenario to obtain an extended field range. 
 We discuss in some
 detail the case of a wrapped $D$5-brane, where inflation naturally ends as the cycle the
 brane wraps shrinks to zero size. We then look at DBI inflation in this new set-up. We show that
 we can match the Cosmic Microwave Background (CMB) data in a regime where we expect the 
 supergravity description to be valid. On the other hand we show that as the brane moves down the throat, the backreaction rapidly becomes important. We then conclude with possible resolutions and a general discussion.
 
\section{Preliminaries}
We limit ourselves to string theory vacua which admit a low-energy
description at the inflationary scale and where the compactified
dimensions are large compared to the string scale. We consider a warped compactification of
string theory with a metric
\be \label{10Dwarp} ds_{10}^2 =
h^{-1/2}(y) g_{\mu \nu}\,dx^{\mu} dx^{\nu}+ h^{1/2}(y)
g_{mn}dy^mdy^n\; ,
\ee
where the warp factor $h$ depends only on the coordinates $y$ of the extra dimensions.
In these constructions the four dimensional effective Planck mass can be
calculated in terms of the warped volume $V_6^w$ of the compact manifold
\ba
\label{mplanck} 
M_p^2 = \frac{V_6^w}{\kappa_{10}^2}, 
\ea 
where
$\kappa_{10}^2 = \hf (2\pi)^7 g_s^2 \ap^4$ is the ten dimensional
gravitational constant (dependent on the string coupling $g_s$ and the string scale $\sqrt{\ap}$) and $V_6^w = \int d^6y \sqrt{g} h(y)$ \cite{DeWolfe:2002nn}.

The inflaton field $\phi$ can be a closed string modulus, an open string modulus (the
position or deformation
 of branes - see \cite{Dutta:2007cr} for an example of the latter) or any combination of those.  We want to discuss the maximal displacement of the field in Planckian units
\be
\frac{\Delta \phi}{M_p}\; ,
\ee
such that the low-energy description is still consistent.
Note that this is really a computational limit and it is entirely
possible that the inflaton could have an unbounded range in some stringy models
that go beyond the supergravity approximation.

For a single field model the maximum range of the inflaton field is
determined by the size of the moduli space of that field. If the
moduli space is compact, then the inflaton field must have a limited
range. Some typical examples of inflaton fields with a compact moduli space are the
axion (with a circle as moduli space $S^1$) or the position of a
$D$3-brane filling out the 4D external space-time
whose moduli space is the full internal Calabi-Yau. An example of a non-compact inflaton field could be the
volume modulus which can go all the way to infinity in the
decompactification limit. 

\subsection{Tensor Modes, Field Range and the Lyth Bound}

In addition to constraining the amount of inflation, the
field range determines how much tensor signal one should expect for
a given model. In slow-roll inflation, one can directly relate the
variation of the inflaton in terms of e-folds $\N$ to the tensor-scalar
ratio $r$
 \be 
 \frac{1}{M_p}\frac{d\phi}{d\N}=\sqrt{\frac{r}{8}}\; .
\ee 
This relation is generically modified for a general kinetic term
although for the case of DBI (the square root), it is
unchanged \cite{Baumann:2006cd}. One can rewrite the above equation in terms of the tensor/scalar ratio $r_{CMB}$ measured at the CMB pivot scale 
\ba
\label{lythbound} 
\r = \frac{8}{(\Ne)^2}\left(\frac{\Delta\phi}{M_p}\right)^2 \; ,
\ea 
where 
\ba 
\Ne = \int_0^{\NE} d\N\left(\frac{r}{\r}\right)^\hf\; .
\ea 
The parameter $\Ne$ is a measure of how much the tensor scalar ratio changes
 \cite{Easther:2006qu}.
For standard slow-roll inflation, a conservative bound can be put on
this parameter $\Ne \sim 30$ \cite{Baumann:2006cd}. For DBI
inflation, the varying sound speed $c_s\sim1/\gamma$ adds some freedom and it is
reasonable to lower $\Ne$ all the way down to $1/\epsilon\sim10$.

Taking the maximum field range allowed by a given microscopic theory gives the so called Lyth bound \cite{Lyth:1996im} on the tensor-scalar ratio
\begin{align}
r < \frac{8}{(\Ne)^2} \left(\frac{\Delta \phi}{M_p}\right)^2_{MAX}\; .
\end{align}
Generally, one would expect that $\frac{\Delta\phi}{M_p} \ll 1$ in
order for the supergravity approximation to be valid and that would
mean that $r$ is very small. 

\subsection{Field Range in $D$-brane Inflation}

Let us first discuss the field range allowed for the
position/deformation of a $D$-brane where the moduli space is the
compact manifold itself (or a submanifold thereof). 
In the following we would like to show how the field range
increases if the inflaton is described by a wrapped brane instead of a $D$3-brane.

\subsubsection{Brane Inflation on a Torus.}

As a first example let us take a simple compactification of string
theory on a symmetric six dimensional torus with size $L$ and volume
$(2\pi L)^6$ \cite{Burgess:2001fx}. Taking the inflaton field $\phi$
to be related to the coordinate $x$ on the torus by a normalization S
through $\phi = \sqrt{S} x$, the maximal field range is given by the
size of the compactification $x\sim L$ 
\ba 
\left(\frac{\Delta
\phi}{M_p}\right)_{MAX}= \frac{\sqrt{S \pi} g_s l_s^4}{L^2}\; , 
\ea
where we have defined $l_s = \sqrt{\ap}$ and we have used the
definition of $M_p$ in Eq. (\ref{mplanck}). For a $D$3-brane with
normalization $S = T_3 = 1/((2\pi)^3 g_s l_s^4)$ we get
\ba
\left(\frac{\Delta \phi}{M_p}\right)_{MAX-D3}= \frac{\sqrt{g_s}
l_s^2}{2\sqrt{2}\pi L^2}\; . 
\ea
In a valid supergravity approximation $L > l_s$ and $g_s <1$, which
means that for a $D$3-brane the field range is always small. 
In more general cases, the relation
between the inflaton and the coordinates can be more complex but we will
stick to this simple constant of proportionality in this paper. It is a good approximation
to more realistic cases.  

One can
possibly extend this field range in several ways. First one can imagine using a
stack of $n$ branes and then $\phi$ is a collective field for the
position of all these branes (like in \cite{Cline:2005ty, Thomas:2007sj} or, in a
different context, \cite{Becker:2005sg, Krause:2007jr}, \cite{Dimopoulos:2005ac, Easther:2005zr})
or \cite{Singh:2006yy, Panigrahi:2007sq}.
This would lead to a  $\sqrt{n}$ factor
enhancement of the field range. The value of $n$ is, however,
limited by the fact that the backreaction of the brane stack must be
kept under control.

Another possibility is to move in a diagonal of the torus. Then the
inflaton is a collective field made of various coordinate
directions. This type of behavior only enhances the field range by
a factor proportional to the square root of the number of (real)
dimensions of the moduli space. For the position of a $D$3-brane
this is just six. 

Both previous ideas involve using a collective coordinate
and only lead to a numerical enhancement of the field range.
A more dramatic possibility is to
consider a different normalization by using different branes. Indeed
if one uses a wrapped $p$-brane (with $p>3$) the normalization would
generically depend on the size of the torus $ S = (2\pi L)^{p-3}
T_{p+1}$ and one finds for a $D$5 and $D$7 respectively \ba
\left(\frac{\Delta \phi}{M_p}\right)_{MAX-D5} &= &\frac{\sqrt{g_s} l_s}{2\sqrt{2}\pi L}\; ,\\
\left(\frac{\Delta \phi}{M_p}\right)_{MAX-D7} & = &\frac{\sqrt{g_s}
}{2\sqrt{2}\pi}\; . \ea Note that these quantities are always smaller than one
 in a consistent supergravity approximation \footnote{The case of the $D$8 brane
 is particularly interesting since it seems that one could get a parametrically large
 field range proportional to $L$.  This is incorrect though and
 we expect in this case that the backreaction will be important
 and reduce the field range to value parametrically smaller than 1.
 We thank Juan Maldacena for pointing this out to us.}.  It is important to note that
 backreaction of the brane on the geometry will change the volume of the internal
 space.

\subsubsection{Brane Inflation in a Warped Throat}

In KKLMMT \cite{Kachru:2003sx} it was argued that inflating in a warped throat flattens the potential and this geometry has since been widely used in the context of string cosmology. 
Most of the current models are based on the Klebanov-Strassler (KS) throat \cite{Klebanov:2000hb}
but we emphasize that the exact metric is not necessary
for many calculations. Instead, up to details of brane embeddings, it is enough
to know the warp factor for a first pass on the cosmology. A general discussion of accelerated expansion in these kind of geometries can be found in \cite{Neupane:2006in}. 
If the space is $AdS_5\times X_5$ 
the metric may be written as
\be 
\label{10Dwarp} 
ds_{10}^2 =
h^{-1/2}(\rho) g_{\mu \nu}\,dx^{\mu} dx^{\nu}+ h^{1/2}(\rho)
\left(d\rho^{2} + \rho^{2} ds_{X_5}^2 \right) \; ,
\ee 
where the warp
factor is $h(\rho) = R^4/\rho^4$, $\rho$ is the radial coordinate (r
is used for the tensor-scalar ratio), $g_{\mu\nu}$ is the Friedmann-Robertson-Walker metric, and the AdS scale $R$ is given by
 \ba
R^4 = \frac{4\pi g_s N\pi^3 \ap^2}{v} \; .
\ea 
Here $N$ is the
amount of $F_5$ flux ($D$3-brane charge) on the cone and $v = \rm{Vol}(X_5)$ is
the volume of the base. In order for
the internal space to be finite, this throat must be glued to a
compact manifold of finite volume.  The gluing can be done at a
distance $\rho_{UV} \sim R$ from the tip of the throat where
$h(\rho_{UV}) = 1$.

Now assuming that most of the volume is coming from the warped
throat and that the bulk contribution is subdominant we get the
following six dimensional volume
\begin{align}
V_6^w \sim  V_6^{\rm{throat}}& = \int_0^{\rho_{UV}} d\rho\; \rho^5 h(\rho) \int d\Omega_{X_5}\; ,\nonumber\\
& = 2\pi^4 g_s N \ap^2 \rho_{UV}^2\; .
\end{align}
Taking the maximum range for the inflaton to be $(\Delta\phi)^2 = S
\rho_{\rm{UV}}^2$ together with the value of $M_p$ from
Eq.(\ref{mplanck}) we get an upper bound for the inflaton variation
\ba\label{D3fieldrange} \left(\frac{\Delta\phi}{M_p}\right)_{MAX} <
2\sqrt{\frac{\T}{N}} \ea where $\T = S (2\pi)^3g_s\ap^2 =
\frac{S}{T_3}$ and $T_3$ is the $D$3-brane tension. So for a
$D$3-brane where $S = T_3$ we see that the range is at its
maximum $2/\sqrt{N}$. In order to have a valid supergravity
description one usually requires $N> 10^4$ (then $R > l_s$ for $g_s \sim 0.1$) 
and this means that the
field range for a $D$3-brane in these throats is usually much
smaller than Planck scale. Including the bulk one might expect
that it is possible to achieve a bigger field range, but of course $M_p$ increases with the bulk volume.

Allowing movement in the angular directions extends the field range slightly (with the inflaton $\phi$ now being a collective field, a combination of the radial and angular coordinates). However, recent work in the subject \cite{DeWolfe:2007hd} (see also \cite{Easson:2007fz}) shows that  this is usually not very helpful and gives only a numerical factor enhancement. 

For a given background, the one potentially free parameter in the calculation above is the normalization of the inflaton field. Instead of using a $D$3-brane, we now consider the inflaton as the position of a wrapped brane in the throat.

\subsection{Wrapped Branes in the Throat}

Imagine a $D$5  \footnote{Most of the work in this paper can be simply generalized to any $(p,q)$ 5-brane.}
brane wrapping a 2-cycle inside of $X_5$.
We then expect the normalization to be of order $S \sim p R^2
T_5$ where $R$ is the AdS scale, the only length scale relevant
to this problem and $p$ is the winding number (see the next section
for more details). From this we find that 
\ba
 \T = \frac{S}{T_3} \sim \frac{ p R^2}{(2\pi)^2\ap} \sim p \left(\frac{ g_s N}{v}\right)^\hf\; .
\ea 
Taking the inflaton to be related the brane position as before
$\phi_{max} = \sqrt{\T} \rho_{UV}$ and $\phi_{min}\approx0$, we find
 \ba
\left(\frac{\Delta\phi}{M_p}\right)_{MAX-D5} < 2\sqrt{\frac{\T}{N}}
\sim 2 p^\hf\left(\frac{g_s}{N v}\right)^{1/4}\; .
 \ea
This is a rather interesting result. In addition to offering new
degrees of freedom to vary we see
that the dependence on the $D$3-brane charge $N$ now enters as the $-1/4$ power. Then for a given value of the background charge the field range will be larger for a wrapped $D$5-brane than for a $D$3-brane.

The situation is even better for a $D$7-brane wrapping a 4-cycle in $X_5$.
The maximum field range is
\ba \left(\frac{\Delta\phi}{M_p}\right)_{MAX-D7} <  2\sqrt{\frac{\T}{N}}
\sim \left(\frac{p
g_s}{v}\right)^{1/2}\; , \ea
and this is completely independent of $N$!

Since wrapped branes have a naturally larger field range,
using these objects for the inflaton may provide inflationary models that can match cosmological data.
To examine this claim we will
focus in the rest of this paper on the $D$5-brane case and we will concentrate on
the DBI type of inflation with a quadratic potential. We do this to compare with the well-studied $D$3 case. 
As we will show, the problems found in \cite{Baumann:2006cd} are essentially resolved for a $D$5 brane inflaton. Nevertheless, the situation is not perfect. In particular, the backreaction for a $D5$-brane is more important than for a $D3$-brane. We expect (dimensionally) the $D$7 brane to have an even larger backreaction and for this reason we will not consider it much further in this paper. 

In the most common model of brane inflation one typically considers
a brane and an anti-brane, with the inflaton the distance between the branes. Inflation ends when the branes approach within a string length, a tachyon develops and the branes annihilate.
Phenomenologically this is hybrid inflation, although at a string length distance the effective
field theory breaks down and one must include higher
open string modes living on the branes. The end of inflation
in these scenarios is then very stringy and interesting physics comes out. The reheating era starts with a phase dominated by massive non-relativistic closed string modes \cite{Lambert:2003zr}
and $(p,q)$-strings of cosmic size are
produced \cite{Jones:2002cv, Sarangi:2002yt}. These could potentially be
observed today as cosmic strings.

If one considers inflation from a wrapped $D$5-brane in a type
IIB vacuum of the KKLT type \cite{Kachru:2003aw},
(i.e. a vacuum with $D$3-branes and/or $D$7-branes)
then there is no need of an anti-brane to break supersymmetry (see \cite{Buchel:2004qg} from some early work on the subject). The $D$5-brane itself, when extended in the three non-compact spatial directions, preserves a different
supersymmetry from the rest of the background. This means that the $D$5-brane is a
non-BPS object which is unstable and can ``self annihilate".
This interesting exit scenario resembles what
happens in a geometric transition where a $D$-brane is transformed into flux.

One can easily see this instability in the warped deformed conifold geometry
since there is no non-trivial 2-cycle in this case (the second Betti number vanishes)
and it is always possible
for a wrapped $D$5-brane to shrink to a point. When the radius of the brane
reaches string scale, a tachyon develops on the brane and it``self-annhilates".
Alternatively, for a throat 
with $M$ units of $F_3$ we can imagine forming a small $D$5-brane (wrapping a 2-cycle
$p$ times) at the bottom of 
the throat and pulling it out to some fixed value of $\rho$. 
Inside the $D$5-brane bubble, there are $M-p$ units of $F_3$ flux while outside the original $M$ units $F_3$ remain (see Fig. 1). If we then let the $D$5-brane go, it will shrink back 
 and self-annihilate leaving the background charge $F_3$ unchanged (at $M$).
\begin{figure}[ht]
\label{cartesianD5}
\centering
\includegraphics[width=8cm]{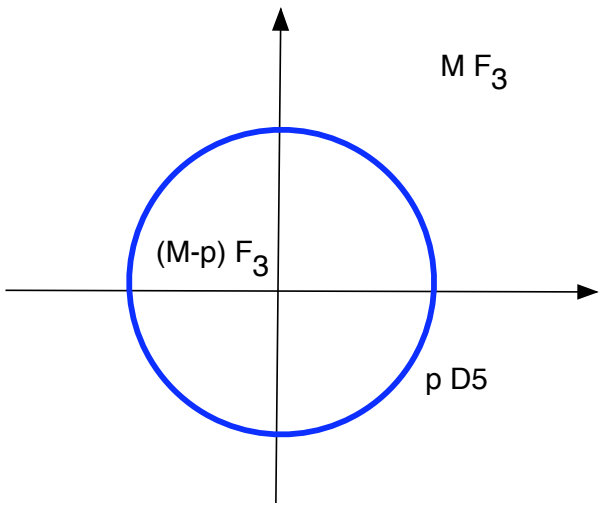}
\caption{A $D$5-brane wrapping a two-cycle $p$ times that shrinks to a point.  We are showing here the system in Cartesian coordinates where the $D$5 is at a fixed radius $\rho_0$. Below $\rho_0$, there are $M-p$ units of $F_3$  while above it the original background $M$ units remain.  When the radius reaches the (accordingly warped) string scale, a tachyon develops on the brane. One can think of this as brane-antibrane annihilation between antipodal part of the brane on the $S^2$ it wraps.}
\end{figure}

Whether one can naturally start this scenario and get inflation as the $D$5-brane shrinks to zero size will depend on the details of the potential the $D$5-brane experiences. This is a model dependent question and it would be interesting to calculate precisely what the potential for a $D$5-brane is
in a particular compactification. In general one might expect many different
metastable vacua. Indeed in the decompactification limit such vacua were constructed in e.g.
\cite{Aganagic:2006ex}. Such metastable vacua could offer a nice starting point for inflation.
One can imagine that the $D$5-brane wrapping an $S^2$ of the Calabi-Yau tunnels out and
falls into a throat.
In this paper, we limit ourselves to scenarios where sufficient e-folds of inflation occur while the
$D$5-brane falls down the throat.  

\subsection{Specific Implementation in the Warped Deformed Conifold}

The warped deformed conifold or Klebanov-Strassler (KS) throat is of
the general form given in Eq.(\ref{10Dwarp}) where the base approaches
$T^{1,1}$ in the UV with the following metric
\ba 
\label{6t11}
ds_{T^{1,1}}^2 &=\frac{1}{9}(\,d\psi +\sum_{i=1}^{2}
\cos \theta_i\, d \phi_i\, )^2 \\
&+ \frac{1}{6}\sum_{i=1}^{2}
 (\, d\theta_i^2 +  \sin^2 \theta_i\, d\phi_i^2\,)\, ,  \nonumber
\ea 
Here ($\theta_{1}, \phi_1$) and ($\theta_2, \phi_{2}$) are coordinates on each $S^2$ and $\psi$ is the coordinate of a $U(1)$ fiber over
the two spheres. This space has a $\mathbb{Z}_2$ symmetry
exchanging the two 2-spheres.  In the IR, the KS throat reduces to
$S^3\times \mathbb{R}^3$.  One $S^2$ shrinks to zero size
while the other joins with the fiber to give an $S^3$. In addition
the KS throat has the following background fluxes
 \ba
 \label{fluxes}
B_2 &= &\frac{3g_s M \ap}{2}\ln (\rho/\rho_0) \omega_2\nonumber\, ,\\
dC_2 & = &\frac{M\ap}{2}\omega_3\nonumber\, ,\\
dC_4 & = &\mathcal{F}_5 + \star\mathcal{F}_5-B_2\wedge F_3\nonumber\, , \\
\mathcal{F}_5 &=&\frac{h^2r^5\alpha^{\prime2}}{54g_s}\partial_r(h^{-1})\omega_2\wedge\omega_3\nonumber\\
&\approx& 8\pi^4 N \alpha^{\prime2}\frac{\omega_2\wedge\omega_3}{27v}
\ea
where $N$ and $M$ are the number of $D$3-branes and fractional $D$3-branes respectively,
$\omega_2 = \hf(\sin\theta_1d\theta_1\wedge d\phi_1 - \sin\theta_2d\theta_2\wedge d\phi_2)$
and $\star \omega_3 = \frac{1}{\rho h}\omega_2\w d\rho\w dx^0\w\cdots dx^3$. For $\mathcal{F}_5$ we have
given the leading term
only. Finally, in the expression for $B_2$, $\rho_0$ is chosen to be the IR scale such
that $B_2$ vanishes a the tip of the throat.
For a nice review of this construction, see \cite{Herzog:2001xk}.

\subsubsection{$D$5-Brane Embedding}

We want to look at a $D$5-brane in this background at some radial
position $\phi = \rho \sqrt{S}$ for some normalization $S$. The $D$5-brane
 extends in all the 3+1 non-compact directions (hence can act
as a vacuum energy) and wraps a 2-cycle in $T^{1,1}$.  This is not a supersymmetric
embedding as the $D$3-$D$5 brane system breaks supersymmetry unless the branes share only
two spatial directions.

We assume
that this embedding is fixed in the angular direction
 and we will not consider fluctuations
around it.  One can generally imagine various embeddings and while
details of the physics will depend on which one is chosen to start
with, the leading term appears to be insensitive.
A general discussion of various possible
 embeddings of a $D$5-brane in this kind of geometry can be found in \cite{Canoura:2005uz, Arean:2004mm}.

 Following this discussion we take the worldvolume coordinates to be
\be
\xi^\mu = (x^0, x^1, x^2, x^3, \theta_1, \phi_1)\; .
\ee
 There exists a natural 2-cycle in $T^{1,1}$ where we take
  $\rho$ and $\psi$ to be constant, $\theta_2 = f(\theta_1)$ and $\phi_2 = g(\phi_1)$.
For this type of embedding, the pullback metric on the $D$5-brane is diagonal and is given by
(spatial
derivatives are dropped since during the inflationary era these
terms are highly suppressed by the scale factor)
\begin{eqnarray}
g_{00} & = &-h^{-1/2} + h^{1/2} \dot\phi^2/S\; ,\\
g_{ii}  &= &a(t)^2 h^{-1/2}\nonumber\; ,\\
g_{\theta_1,\theta_1} & = &\frac{h^{1/2}\phi^2}{6S} (1 + f^\prime(\t_1)^2)\; ,\nonumber\\
g_{\phi_1,\phi_1} & = & \frac{h^{1/2}\phi^2}{S} X(\t_1,\phi_1)\; ,\nonumber
\end{eqnarray}
where $i$ runs from 1 to 3, we have introduced the FRW scale factor $a(t)$ and
\bas
&X(\t_1,\phi_1)  = \frac19\cos^2\t_1 + \frac16\sin^2\t_1 +  \\
&\left(\frac19\cos^2\t_2 + \frac16\sin^2\t_2\right)g^\prime(\phi_1)^2
+ \frac29\cos\t_1\cos\t_2g^\prime(\phi_1)\; .
\eas

The pullback for the NS-NS tensor is
\ba
P[B_2] &= &\frac{3g_s \ap M\ln(\rho/\rho_0)}{4}\left(\sin\t_1\right.\nonumber \\
&&\left. - \sin(f(\t_1))f^\prime g^\prime\right) d\t_1\wedge d\phi_1\; .
\ea
 More specifically, we can use the following embedding
$\theta_2=-\theta_1$ and $\phi_2=-\phi_1$
 (which is the same as the coordinate identification that shows the reduction of the
 metric to an $S^3$ in the IR, that is to say that this is the 2-cycle that shrinks to
 zero size in the KS throat). 
This set-up is shown in Fig. (2).  For this particular embedding, 
\ba
X(\t_1,\phi_1) &= & \frac13\sin^2\t_1 \; ,\nonumber\\
P[B_2] &= &\frac{3g_s\ap M\ln(\rho/\rho_0)}{2}\sin\t_1 d\t_1\wedge d\phi_1\; .\ea
 \begin{figure}[ht]
 \label{rdirection}
\centering
\includegraphics[width=8cm, height= 8cm]{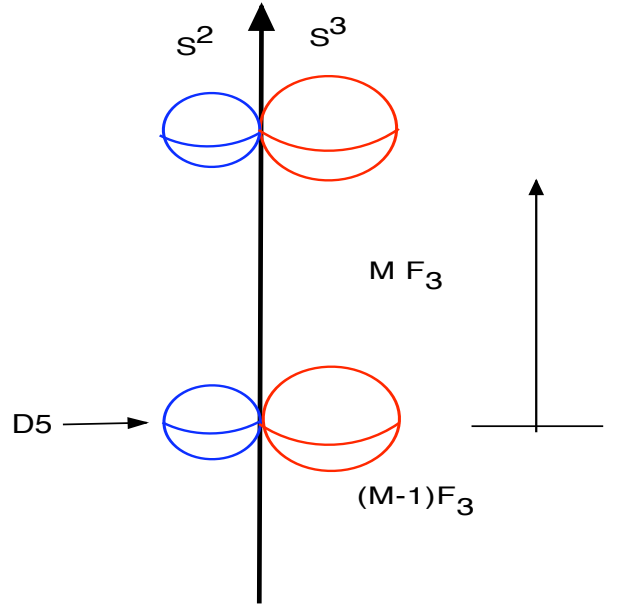}
\caption{We show the $\rho$ direction together with the $S^2$ and $S^3$.  There are $M$ units of $F_3$ threading through the $S^3$.  We wrap the $D$5-brane on the $S^2$ at some fixed position on the $\rho$ axis. Below that value of $\rho$ (below the dark line in the picture) there is one less unit of $F_3$ threading through the $S^3$.}
\end{figure}

\subsubsection{The DBI Action}

Considering the case with no gauge flux on the worldvolume, the following pullback of the metric and NS-NS tensor
 \ba
\sqrt{-\det(P[g+B]} & = &h^{-1} a(t)^3 \sin\t_1 \left[ \left(1-
\frac{h\dot\phi^2}{S} \right)
\mathcal{F}(\phi)\right]^{1/2}\nonumber\\
\mathcal{F}(\phi)   &= &\frac{h\phi^4}{9S^2} + \frac{9
g_s^2\ap^2 M^2}{4}(\ln(\phi/\phi_0))^2\; . \nonumber
\ea 
Note that different choices of
embeddings on the $D$5-brane would
affect $\mathcal{F}$ through the log part.  
The NS-NS tensor is a decreasing function of 
$\phi$. The maximal value is obtained for $\phi_{edge} = \sqrt{S}R$,
\ba\label{neglectB2}
\frac{9g_s^2\ap^2M^2}{4}(\ln(\phi_{edge}/\phi_0))^2 \approx \pi^2\ap^2K^2
\ea
where $\phi_0 \approx \sqrt{S} (g_sM\ap)^{1/2} e^{\frac{-2\pi K}{3Mg_s}}$ and $K = N/M$.
In the AdS part of the throat $h \propto R^4/r^4 \sim R^4
S^2/\phi^4$.  For $K < 2\pi\sqrt{\frac{g_s N}{v}}$, we can neglect
the contribution coming from $B_2$ and 
\ba 
\mathcal{F} \approx R^4/9\; . 
\ea
Since
$\mathcal{F}$ is roughly constant in the AdS part of the throat the choice of embedding and gauge flux is only marginally
important.
Note that if $q$ units of magnetic flux on the brane worldvolume are turned on, there would be an additional term going like $\ap^2 q^2$ in $\mathcal{F}$.  
Now putting all this in the DBI action for the $D$5-brane
\ba S_{D5}& = - T_5 \int d\t_1 d\phi_1 d^4 x \sin\t_1
\frac{a(t)^3h^{-1}R^2}{3}
\sqrt{1-h\dot\phi^2/S}\; ,\nonumber\\
& = - \int d^4x \frac{a(t)^3 h^{-1} 4\pi R^2T_5}{3}\sqrt{1-h\dot\phi^2/S}\; ,\nonumber\\
& = - \int d^4x a(t)^3 f(\phi)^{-1} \sqrt{1-f(\phi) \dot\phi^2}\; .
\ea
where in the last line $\phi$ is now canonically normalized with $ S = \frac{4\pi R^2T_5}{3}$ and
 $f(\phi) = \frac{3 h(\phi)}{4\pi T_5 R^2 }$.  This is exactly the kind of normalization we used in
 section (II).

One can also consider the possibility of having a winding number $p$ as well as the possibility of
wrapping a base that is orbifolded.  In the case of the orbifold, the base will have a smaller volume
and the two cycle wrapped by the $D$5 might or might not be orbifolded.  To account for that we introduce a new
parameter $a$ (not to be confused with the scale factor) that determines the effect of the orbifolding on the 2-cycle's volume
\ba
\int_{S^2} d\Omega_2 \rightarrow \frac{1}{a}\int_{S^2} d\Omega_2\, .
\ea

So a $Z_n$ orbifold of $T^{1,1}$ in the $\psi$ direction
would give $v = v_{T^{1,1}}/n$ with $a = 1$ while orbifolding $\theta_1$ would give $a = n$ (in addition to the overall
change in the base volume).  Hence the final result for the
normalization of a $D$5-brane with winding number $p$ in the KS throat with a possible
orbifolding of the base is \ba S = \frac{4\pi R^2}{3}\frac{p}{a}
T_5 \; , \ea and the field range is bounded by (using Eq. \ref{D3fieldrange}) 
\ba
\label{fieldrangemax}
\left(\frac{\Delta\phi}{M_p}\right)^2 & \leq & \frac{2^3\pi}{3}
\frac{p}{a}
 \left(\frac{g_s}{Nv}\right)^{\hf}\; .
\ea

We note that in this paper, we will take $p=a=1$ for most cases while we will find 
that we usually need some orbifolding $n$.

\subsubsection{The Chern-Simons Action}

The Chern-Simons (CS) part of the $D$5-brane action has two non-vanishing
terms in the KS throat \ba S_{CS} = \mu_5 \int P[C_6 + C_4\w(B_2 +
2\pi\ap F)]\; . \ea The terms proportional to $C_2$ and $C_0$ vanish
because they have no support in $\mathbb{R}^{3,1}$ with no fluxes
turned on in these directions.

$C_6 $ is Hodge dual to $C_2$ with $dC_6 = \star dC_2$. In the KS
throat we have $dC_2 = \frac{M\ap}{2}\omega_3$ and from this we
obtain \ba
C_6 &= &\frac{3M\ap}{2} g(\rho) dx^0\cdots \w dx^3 \w \omega_2 \;,\nonumber\\
g^\prime(\rho) &= &\frac{1}{\rho h} \; .
\ea 
Integrating $g^\prime(\rho)$
with the AdS warp factor we get
 \ba 
 g(\rho) = \frac{1}{4h} + c \; .
 \ea
Requiring that the flux vanishes as $\rho \rightarrow 0$ we can set
$c=0$. Similarly, $C_4$ is given in the KS throat by 
\ba C_4 =
\frac{1}{g_s h} dx^0\cdots \w dx^3 + \cdots \; ,
\ea 
where we ommitted
terms with no support on $\mathbb{R}^{3,1}$. Taking $B_2$ given by Eq.(\ref{fluxes}) together with
$q$ units of $F$ flux ($F =
\frac{q}{2}\sin\theta_1d\theta_1d\phi_1$), doing the pullback and
integrating over the $S^2$ coordinate, we get the following
contribution to the $D$5-brane action
 \ba 
 \mu_5 2\pi \ap \int d^4x
\left(\frac{3M}{4 h} + \frac{1}{g_s h}\left(3g_s M\ln (\phi/\phi_0)
+ 2\pi q\right)\right) 
\ea 
This goes like $\phi^4
(1+\ln\phi)$ and so we find that the CS term provides a
quartic piece to the potential of a $D$5 in this background.
This is just one piece of the potential; many other effects could contribute terms and we 
will come back to this issue later on. Also, for the rest of this paper, we consider
the case with no flux on the brane $q=0$ although we will comment on
its effect in the conclusion.

\section{DBI Inflation from Wrapped Branes}

We now consider the inflationary predictions of this model. The full potential for a $D$5-brane in a KKLT like setting has not been calculated and we expect other terms
will be generated once moduli stabilization effects are taking into account.
We do not expect the resulting potential to be generically flat enough for slow-roll, although it
might be tunable to be so. For easy comparison to the current literature on $D$3 inflation we will assume that the potential is not flat and instead dominated by a quadratic piece. We will return to a discussion of the potential, including the quartic term from the Chern-Simons action in the next section.

\subsection{General Constraints}

Let us first take a simplified look at the fit to data. As we will
see, the wrapped $D$5-brane system works with a smaller background charge $N$ compared to a $D$3-brane. We note that the changes are not minor, as the fit to the data depends on
different \emph{powers} of $N$ and thus we find very different
results than in the previous analysis for the $D$3-brane case.

First, we will consider the most general case of any wrapped brane
in a throat with an asymptotic AdS regime such that $f(\phi) = S^{-1}
h(\phi) = \frac{R^4 S}{\phi^4}$. We take the potential to be
quadratic $V \sim \hf m^2\phi^2$ and $\phi$ to be decreasing (the brane is moving into
the throat).  Such a potential gives rise to a Hubble constant that is linear in $\phi$ from the FRW equation
$H^2 = \frac{V}{3M_p^2}$.  This is usually referred to UV DBI and we will have some words later on
the opposite regime where the brane is going out of the throat.  The following result 
is useful in deriving simple relations  \cite{Silverstein:2003hf}:
\ba
\gamma = \frac{1}{\sqrt{1 - f(\phi)\dot\phi^2}} = \sqrt{1+ 4M_p^4 f(\phi)H^{\prime2}}\; .
\ea
We can define three dimensionless
constants in this problem \ba
A &\equiv& H^\prime = \frac{m}{\sqrt{6} M_p}\; ,\nonumber\\
B &\equiv& \frac{\phi}{M_p}\; ,\nonumber\\
C &\equiv & R^4 S\; . \ea Note that $B$ is just the field range while
$C$ encodes the information from the normalization of the field. For
a $Dp$ brane (with $p>3$) we expect $C \propto N^{(p+1)/4}$. We will
be in the DBI regime when $\gamma \sim 2M_p^2 f(\phi)^{1/2} H^\prime >
1$.
 We can fix these three constants by
using three pieces of data. The strongest observational constraints come from the bounds on non-Gaussianity ($|f_{NL}| < 300$) \cite{Creminelli:2006rz},
on the tensor-scalar ratio $r<0.3$ and from the power spectrum normalization $P_S
\sim 2\times10^{-9}$ \cite{Spergel:2006hy}. We will not worry about the exact value of the spectral index here; $n_s\rightarrow1$ in the DBI limit of this system, but one can numerically analyze the ``intermediate regime" to more accurately match observations \cite{Shandera:2006ax, Bean:2007hc}. Since $f_{NL} \propto \gamma^2$ the bound on non-Gaussianity translates directly into a bound on the Lorentz
factor at CMB scales $\gamma < 30$.  We
can express these three observables in terms of the dimensionless
constants above
\ba 
\label{dataconstraints}
\gamma & \sim &2M_p^2 f(\phi)^{1/2} H^\prime = \frac{2\sqrt{C} A}{B^2} < 30\; ,\\\nonumber 
r&=&\frac{32 M_p^2}{\gamma^2\phi^2} = \frac{8B^2}{CA^2} < 0.3\; ,\\\nonumber
P_S&=&\frac{H^4\gamma^2}{16\pi^2M_p^4H^{\prime2}} = \frac{A^4C}{4\pi^2} \sim 2 \times 10^{-9}\; . 
\ea 
From the second equation we have that
\ba
B^2 &= &\left(\frac{\Delta \phi}{M_p}\right)^2 = \frac{32}{r\gamma^2}\label{B}\; ,\\\nonumber
& \geq & \frac{1}{9}\; .
\ea 
This is a very conservative bound
since it takes $r$ and $\gamma$ to be both at their maximum at
the same time.  As one can see, DBI inflation demands nearly Planckian field range
in order to work. 

Now using the two remaining equations one can find the constraints on $A$ and $C$ to
be
\ba\label{ACbound}
A &< &2\times 10^{-4}\label{AC}\;,\\\nonumber
C &> & 10^{8}\; .
\ea
The constraint on $A$ is
something that in the absence of calculating all the contributions
to the $D$5-brane potential in a given compactification, we must
assume is possible to achieve. It says that the mass of the inflaton
field should be at least four orders of magnitude below the Planck
scale which is sensible. (Of course, $A$ too small moves the system out of the DBI regime and toward slow-roll.) One can similarly find a lower bound on $A$ and an upper bound on $C$ by taking $B^2<1$. Then $A>2\times10^{-5}$ and $C<6\times10^{11}$. But since we know the microscopic limits on the field range here, we can instead use that to examine the constraints in Eq.(\ref{ACbound}) in more detail.

For DBI inflation with a $D$3-brane, $C = R^4 T_3 = \frac{\pi
N}{2v}$. Taking this into account, the conditions in Eq.(\ref{B})
and Eq.(\ref{AC}) read (taking $B$ to be the maximum value allowed Eq.(\ref{D3fieldrange}))
\ba
\label{boundsD3}
\frac{N}{v} &> & 8 \times 10^7\\\nonumber
N &< &36
\ea 
which can only be achieved with a very small volume for
the base but even then the necessarily small value for $N$ forces us into a regime where the supergravity approximation is not trustable.
 On the other hand for a $D$5 wrapping $p$ times on a cycle
 (with orbifolding $a$), $C = \frac{\pi^2}{3} \frac{p}{a}\left(\frac{N}{v}\right)^{3/2} g_s^{1/2}$ and
 the condition in Eq.(\ref{AC}) gives
\ba \label{boundsD5-1}
\frac{N}{v} > 10^5 g_s^{-1/3}\left(\frac{a}{p}\right)^{2/3} 
\ea 
while the condition in Eq.(\ref{B})
(using Eq.(\ref{fieldrangemax}) for the maximal field range) gives
\ba
\label{boundsD5} 
N v <  6 \times 10^3 \left(\frac{p}{a}\right)^2 g_s 
\ea 
As we
said before we now have several new parameters to vary and there
might now be some room in the parameter space that can fit the
data. While these two constraints (\ref{boundsD5-1}, \ref{boundsD5}) are necessary to match data,
 they are not sufficient. For a given data point that satisfies (\ref{boundsD5-1}, \ref{boundsD5}), 
 we still need to make sure
that we solve all of the equations (\ref{dataconstraints}). We leave a more detailed study of the parameter space for future work but note that one can now see sensible points that match the data. For example taking $p\sim a$, $g_s
\sim 0.1$, all constraints are satisfied if
\ba
\label{Nvbounds}
N= 10^4\\\nonumber
v = 1/40
\ea
for which the supergravity solution is expected to be valid. For these values of the parameters we find
\ba\label{results}
\gamma &= &25\; ,\\\nonumber
r & = &0.29\; ,\\\nonumber
P_S & = & 2 \times 10^{-9}\; .
\ea
If we take the background to be an orbifold of $T^{1,1}$ and if we do not orbifold the 2-cycle the 5-brane is wrapping ($a =1 = p$), then we need to orbifold the 
transverse space to it by an amount found from
\ba
v = \frac{16\pi^3}{27 n} = \frac{1}{40}
\ea
which gives $n \approx 735$. This could be relaxed somewhat by allowing for some winding number $p > a$.  We can check also that for this range of parameters, it is self consistent to neglect $B_2$ in Eq. (\ref{neglectB2}). 

Finally, let us mention that two others observables, 
the number of e-folds and the cosmic string
tension, depend on the geometry at the bottom 
of the throat (or at the very least, they depend on the value of $\phi$ at which the throat
is cut off).  A complete analysis for the number of e-folds should include the fact 
that the geometry of the KS throat is modified in the IR  \cite{Kecskemeti:2006cg}. 

The number of e-folds in the AdS part of the throat (and assuming the relativistic solution) is 
\ba
\label{Ne}
N_e & = & \frac{1}{2M_p^2} \int H dt\\
& = & - \frac{1}{2 M_p^2} \int_{\phi_i}^{\phi_f} \frac{H\gamma}{H^\prime}d\phi\nonumber\\
& = & - A\sqrt{C}\ln \left(\frac{\phi_f}{\phi_i}\right)\nonumber
\ea 
where we have used $\gamma = 2A\sqrt{C}\frac{M_p^2}{\phi^2}$ to
get the last line. Taking $\phi_i$ to be the edge of the throat ($\sqrt{S} R$) and $\phi_f$ to be
the tip of the throat then the warping at the bottom is just
 $h_{tip}^{1/4} = \frac{\phi_{i}}{\phi_{f}}$. In order to get more than 60 efolds in the AdS region by itself
 one needs
 \ba h_{tip}^{-1/4} <
e^{-60/(A\sqrt{C})} \; .\ea
We should mention that in such a scenario $\gamma$ increases quite quickly as the branes goes down 
the throat and as we will show, the backreaction becomes important.
For completeness, the cosmic string tension for a (p,q) string in the conifold is given in
  \cite{Firouzjahi:2006vp, Thomas:2006ud, Firouzjahi:2006xa}.
For a $D$-string, this is approximately
  \ba
  G\mu &=& h_{tip}^{-1/2} \frac{G}{2\pi\ap g_s}\, ,
  \ea
where $G$ is Newton's constant. This can easily satisfy the current experimental bound for a small enough warping at the bottom.

\subsection{Bounds on the Tensor-Scalar Ratio}

In \cite{Lidsey:2007gq}
Lidsey and Huston (LH) presented a lower and upper bound 
on the tensor scalar ratio for DBI inflation. The lower bound is particularly
interesting as it is completely independent of the normalization 
of the scalar field.  Indeed under the assumptions that the brane
moves down the throat, that the warp factor decreases monotonically and that the 
the running of non-Gaussianity is smaller than 1, they found \cite{Lidsey:2007gq}
\ba
r > \frac{1-n_s}{\sqrt{f_{NL}}} \sim 0.002\; ,
\ea
which is at the limit of what is expected to be observable (see for e.g. \cite{Bock:2006yf}) .  This bound definitely applies
in our set-up although nothing strictly forces us to have small running. If the running of the non-Gaussianities is large there will be essentially no lower bound.

LH also calculate an upper bound on $r$ from the Lyth bound. This bound
does depend on the normalization and we repeat it here for the $D$5 case. 
 Their approach uses an
even lower bound on $M_p^2$ than what we have used up to now since they consider only a fraction of the volume of  the throat. The advantage of this method is that one finds a very simple expression at the end for the $D$3-brane case.

Conside the variation of the inflaton $\Delta \phi_* $ over the small range of observable efolds where one can treat $r$ as constant. Approximating
the volume of the throat over that range to be  $\Delta V_* \approx \frac{v (\Delta\rho_*)^6}{h_*^4}$, LH find the following maximal field range over that observable range of e-folds
\ba
\left(\frac{\Delta\phi_*}{M_p}\right)^6 < \frac{\kappa_{10}^2 S^2}{f(\phi) M_p^4 v}\; .
\ea
This expression can be written in terms of observables and some background parameters (which cancel out for the $D$3-brane case):
\ba
\left(\frac{\Delta\phi_*}{M_p}\right)^6 < \frac{\pi^3}{16} r^2 P_s^2 \left(1+\frac{1}{3f_{NL}}\right) \frac23\pi\sqrt{\frac{g_sN}{v}} \frac{1}{v} \frac{p}{a}\; .
\ea
Combining this with the Lyth bound, Eq.(\ref{lythbound}), gives
\ba
r_* < \frac{5 \times 10^{-6}}{(\Delta N_{e,*})^6 v} \sqrt{\frac{g_sN}{v}}\frac{p}{a}
\ea
which for our benchmark point of the previous subsection ($p\sim a$, $g_s \sim 0.1$, $v = 1/40$ and $N \sim 10^4$) and  $\Delta N_{e,*} = 1$ gives
\ba
r_* < 0.04\; ,
\ea
which is slightly above the lower bound derived earlier. Again we see that the $D$5 can match the data although without much wiggle room. This result is fairly sensitive to factors of two, and clearly if one allows for some winding number $p$, $r_*$ can be larger. Very interestingly the upper bound is close to the limit of what is expected to be detectable. 
We note that this upper bound is lower then the value we found in (\ref{results}) which can be simply accounted for by having $\Delta N_{e*}$ be slightly less than 1. For comparison, the Lyth bound translate into the following upper bound on $r$ for
the $D$3 brane case: $r < 10^{-7}$ \cite{Lidsey:2007gq}.

\section{Backreaction, KK modes and the Inflaton Potential}
Although the $D$5 system is at this level more successful than the $D$3 system, we now turn to some potential problems.
\subsection{Backreaction}
Unlike the $D$3-brane, the $D$5-brane will have a backreaction on both the warp factor and 
the internal geometry of the base. This makes the full backreaction calculation quite messy, but it is fairly simple to make an order of magnitude estimate. Let us first consider the case of a static (or slowly moving) $D$5-brane. The $D$5-brane is a source
of $F_3$ and so we require $p \ll M$. This ensures that the flux threading the $S^3$ is approximately 
the same below and above the $D$5.

If we integrate over the base, the $D$5-brane looks just like a $D$3 with tension $p T_5 R^2$ from dimensional analysis. The original background of the deformed conifold was made of $N$ $D$3-branes with tension $T_3$.  Demanding
that $p T_5R^2 < NT_3$ gives the following bound
$\frac{p}{2} \ll \sqrt{\frac{Nv}{g_s}}$
which can be satisfied for our benchmark point used in the previous section of $v \sim g_s$ and $N\sim 10^4$ (then $p \ll 10^2$). We see already from this argument that the backreaction of the $D$5 is much more important than for a $D$3.  Furthermore, we neglected the effect on the internal 
geometry. Closer to the $D$5, the metric should approach a $D$5-brane metric (see for example \cite{Polchinski:2000uf} for a nice discussion of this issue).  Nevertheless, this dimensional analysis argument tells us that while the metric will be different close to the $D$5-brane, the overall throat geometry should be left unchanged if the previous condition is satisfied. 

Now all of this analysis was for a static $D$5-brane and should be valid when the brane is moving slowly. If the brane is moving relativistically, the energy of the brane and its backreaction on the metric will be enhanced \footnote{Note that although we assume that the potential energy of the brane dominates the kinetic energy in the $4D$ picture, the potential energy (other than the Chern-Simons term) is not localized on the brane - that is, it originates from the supergravity part of the $10D$ action, not from inside the DBI action. One should check separately that whatever terms contribute to the brane potential do not come from sources which already change the background KS solution. It is an interesting question whether generic large, unwarped potentials are in fact consistent.}.  One can see this by looking more precisely at Einstein's equation for this background. 

From the trace of Enstein's equation \cite{Giddings:2001yu, Giddings:2005ff} and after some manipulations to write the Laplacian of $h$ 
\ba
-\nabla^2h =\frac{\kappa_{10}^2}{2} h^{3/2}(T^{m}_{m} - T^{\mu}_{\mu})^{local} + \rm{fluxes}\; ,\ea
where $\nabla^2$ is the Laplacian operator using the unwarped metric $g_{mn}$ in (\ref{10Dwarp}). We have isolated the local part due to branes. 
Note that the flux part contains power of the warp factor as well as derivative of the warp factor $\partial h$. The original background can be thought of as coming from $N$ $D$3-branes for which the stress energy tensor is 
\ba
\frac{\kappa_{10}^2}{2} h^{3/2}(T^{m}_{m} - T^{\mu}_{\mu}) = 2\kappa_{10}^2N T_3 
\frac{\delta(r)}{r^5}\frac{\delta(\Psi)_{T^{1,1}}}{v}\; .\ea
The $D$3-branes are just a point source in the Green's function for $h$ (see for example \cite{Baumann:2006th}).  For a static $D$5-brane we get that 
\ba
\frac{\kappa_{10}^2}{2}h^{3/2}(T^m_m -T^\mu_\mu) &=& \kappa_{10}^2 p T_5 h^{1/2} \frac{\delta(r)}{r^3}\frac{\delta(\Psi)_{S^3}}{v_{S^3}}\; , \\
& =& \kappa_{10}^2 p T_5 R^2 \frac{\delta(r)}{r^5}\frac{\delta(\Psi)_{S^3}}{v_{S^3}}\; ,
\ea
where $v_{S^3}$ denotes the volume of the transverse $S^3$ to the $D$5-brane. While the angular dependence has changed (from the $\delta$ function), the $r$ dependence remains the same and it looks just like a $D$3-brane with tension $\frac{p T_5 R^2v }{2v_{S^3}}$ (in the following we will take $v/v_{S^3}\propto 1/a$).
Now including the $\gamma$ factor in the stress energy tensor, $T_0^0$ will dominate in the trace, and we get the same answer multiplied by $\gamma/2$.
\ba
\frac{\kappa_{10}^2}{2}h^{3/2}(T^m_m -T^\mu_\mu) & =& \kappa_{10}^2 \frac{\gamma}{2} p T_5 R^2 \frac{\delta(r)}{r^5}\frac{\delta(\Psi)_{S^3}}{v_{S^3}}\; .\ea
 This gives a new bound (now neglecting factors of order 1)
\ba
\frac{p\gamma}{a} \ll \sqrt{\frac{Nv}{g_s}}\; .
\ea
Using the bounds (\ref{boundsD5}) from data we get that 
\ba
\gamma \ll \sqrt{\frac{Nv}{g_s}} \frac{a}{p} \ll 10^{3/2}\; ,
\ea 
in order for the backreaction on the radial dependence of the warp factor to be negligible.
For comparison, the same can be done for the $D$3 case where we have simply that $\gamma T_3 < NT_3$ and using the bounds (\ref{boundsD3}) gives 
$\gamma \ll 36$.
We see that from the backreaction point of view
the $D$5-brane does not fare better than the $D$3. This analysis also shows that while the $D$5-brane can match the data when it is located near the top of the throat, its backreaction is important and cannot be neglected. In particular it is not clear that one can achieve the required amount of inflation. However, it is plausible that the overall cutoff of the throat is not shifted much, and so the approximate calculation of e-folds is still reasonable.

Note that a similar problem should arise if one tries to use $D$7-branes in this set-up. A static $D$7-brane does not perturb the warp factor ($T^m_m - T^\mu_\mu$ vanishes for a static $D$7). This is no longer true if the $D$7 moves relativistically and $T^0_0$ dominates.  It is easy to check that the backreaction is in general bigger for a $D$7 then for a $D$5 in this case.

Let us emphasize again that in addition to all this, the $D$5-brane should perturb the base of the cone and the dilaton (which should now have a weak dependence on the radial position), this is true whether one does slow-roll or DBI inflation with the $D$5.  In order to do a faithful match 
to the data, one should include this backreaction correctly in the calculation. We leave a more detailed analysis of these effects for further work. 

\subsection{The Hubble Scale and KK Modes}

Another quick calculation reveals a similar potential problem with DBI inflation.
One might think that the warping at the bottom of the throat $h_{tip}$ 
can be arbitrarily small but there is an
important consistency check (as pointed out in
\cite{Silverstein:2003hf,  Alishahiha:2004eh}) that the
 Hubble parameter should be less than the mass of the KK modes in order for the truncation to a single-field effective field theory (that we have assumed) to be valid. That is, 
\ba
\label{KK}
m^w_{KK}&>&H\; ,\\\nonumber
\frac{1}{R}\frac{\phi_{IR}}{\phi_{UV}}&>&H_{max}=A\phi_{max}\; ,\\\nonumber
h_{IR}^{-1/4}&>&A\sqrt{S}R^2=A\sqrt{C}\; ,
\ea
since the $h^{(-1/4)}(\phi)=\phi/\phi_{UV}$.
However, consider the number of e-folds (from Eq. (\ref{Ne})):
\ba
N_e &=&A\sqrt{C}\log(h_{IR}^{1/4})\; .
\ea
Combining with the last line of Eq.(\ref{KK}), we find
\be
N_e<-A\sqrt{C}\log(A\sqrt{C})\; .
\ee
This is positive since $A\sqrt{C}$ is less than one (from Eq. (\ref{ACbound})), but it cannot be greater than 1. So it seems that for DBI inflation in an AdS throat, regardless of inflaton normalization, it is impossible to achieve a significant number of e-folds while $m^w_{KK}>H$. 

If the warped KK modes masses are lower than the Hubble scale then three questions follow. First,
is the low-energy description where we discard these massive KK modes still valid during inflation?  
Second, is the mass of these KK modes shifted during inflation to a value of order $H$? Finally, if the modes are significantly shifted, do they affect the inflationary observables? To answer these 
questions properly we would need to know explicitly the 10 dimensional origin of the potential and then ask whether the KK modes ``see" this energy density. The four dimensional effective field theory question goes back to how supersymmetry breaking is mediated to the warped modes at the bottom of the 
throat. In \cite{Burgess:2006mn}, it is argued that the supersymmetry breaking scale can be warped down in this kind of model. Given a more complete description of the set-up (i.e., the source for all terms in the potential) this may be checked in detail.

\subsection{Discussion of the potential}
For the parameter values given in Eq.(\ref{Nvbounds}) and Eq.(\ref{AC}), the Chern-Simons quartic term is 
smaller than the proposed quadratic term as long as the inflaton mass (or parameter $A$) is near its upper bound. However there is a more general question about the potential that arises when $\phi/M_p$ is so close to one - that is, can we be sure that terms depending on higher powers of $\phi$ do not dominate the potential \cite{Easther:2006qu}? This is especially important because if such terms do dominate, the system will most likely not match the data. For example, from Eq.(\ref{infcondition}), we can see that if $H(\phi)=h_n\phi^n$ with $n>4$, the system is not inflating. A stronger version of the observational bound is the generalization of the second line in Eq.(\ref{dataconstraints}), which gives
\ba
\label{strongbound}
r&=&\frac{16\epsilon}{\gamma}=\frac{32n^2}{\gamma}\frac{M_p^2}{\phi^2}\; ,\\\nonumber
\left(\frac{\phi}{M_p}\right)^2&=&\frac{32n^2}{r\gamma^2}\geq\frac{n^2}{9}\; ,
\ea
and so we see $n<3$ is necessary to match the data. Even the case of the quartic term ($H(\phi)=h_2\phi^2$) is a bit tricky, since the system only inflates for $\phi>2/(h_2R^2\sqrt{S})$.

If the potential is dominated by a constant (i.e., hybrid inflation in the usual sense), then the bound above constrains the ratio of coefficients in $H(\phi)$ rather than the field range, which seems more sensible. However, it is easy to show that these models generally have a blue tilt, while CMB data prefer red-tilt. For a thorough investigation of the behavior of the $D$3 system with fairly general $H(\phi)$ and $c_s(\phi)$, see \cite{Peiris:2007gz}.

The IR DBI models \cite{Chen:2004gc, Chen:2005ad}, where branes move out of the throat and $H(\phi)$ is dominated by a constant early on may be better for data fitting at CMB scales. However, these models will be nearly slow-roll soon after CMB scales, and it is not clear where the appropriate potential would come from. If SUSY-breaking effects take place in the bulk, it seems likely that the minimum energy pushes wandering $D$3-branes away from the bulk. However, more imaginative scenarios are probably possible. Certainly for the $D$5 scenario it is difficult to imagine a reasonable scenario that pulls $D$5-branes away from the tip and yet leaves the background solution otherwise stable.

\section{Conclusions}

In this paper, we have shown that one can extend the allowed field range in brane inflation by 
using higher dimensional wrapped branes.  In particular, we presented a model of inflation where
a $D$5 wrapping a trivial two-cycle in the warped deformed conifold travels down the throat until its radius shrinks
to string scale, where it self-annihilates through tachyon condensation. 

Since the field range for the $D$5 is extended compared with the $D$3-brane it is worthwhile 
to revisit the case of DBI inflation with these branes. We have shown that for a quadratic potential the bounds of 
\cite{Baumann:2006cd} are less restrictive and the fit to data can be done. On the downside, we have found that the backreaction is more important for higher dimensional branes and should be carefully checked before declaring this model successful.

Clearly, one needs the actual $D$5-brane potential, which we have not attempted to compute here.  It would be interesting to describe this system through a Kahler/superpotential framework and to work out the explicit dependence of these functions on the $D$5-brane position. Knowing these, one could explicitly compute the potential and check how likely inflation is in this set-up.

It is possible that a slightly different system could be constructed which is less restricted. An interesting possibility is IR DBI, where the brane goes out of the throat with a potential given by $V = V_0 - \frac12m^2\phi^2$. In this case $\gamma$ must start relatively small and decrease, so it might be possible to have the backreaction under control.  We should note however that even in this case, while the backreaction on
the warp factor can be negligible, we still expect a backreaction on the internal geometry of the base and on the dilaton.  Also, one needs to analyse this system at the bottom of the throat where the geometry is very different. 

In this paper we have not included fluxes on the brane. The presence of flux would change the physics of the system significantly since the $D$5-brane would now be charged under $F_5$ and it would no longer self-annihilate as it reaches the bottom. Instead it would stabilize at a fixed radius. Indeed one can think of a $D$5-brane with $q$ units of flux turned on as the dual of the blow-up of $q$ $D$3-branes
\cite{DeWolfe:2004qx}. Anti-$D$3-branes at the bottom would be required to end inflation gracefully in that case. 

This system has been studied already in \cite{Thomas:2007sj, Lidsey:2007gq}.
Let us just point out here that for the flux part to dominate in the DBI action there must be a large number of $D$3-branes in which case there is also a large backreaction. We believe that the simple $D$5 with
no flux will give the smallest backreaction while still enhancing the field range significantly.  Furthermore, the simplicity of the exit of inflation makes this scenario quite attractive.

\emph{ Note added: While this paper was in the final stages of completion \cite{Kobayashi:2007hm} appeared, which contains significant overlap with section II and III.}

{\bf Acknowledgements} We would like to thank Nima Arkani-Hamed, Aaron Bergman, Jason Kumar, Liam McAllister, Rob Myers, Jessie Shelton, Gary Shiu and especially Renata Kallosh, Juan Maldacena and Eva Silverstein for useful discussions. 
M.B. thanks the Harvard physics department for hospitality
and partial support while this work was in progress.
We are grateful to the organizers of the 2007 Simons workshop in mathematics and physics
where this work was finished.
The work of M.B. and L.L was supported by NSF grant PHY-0505757 and the University of Texas A\&M.
The work of S.S. is supported by the DOE under DE-FG02-92ER40699.

\end{document}